\documentstyle[aaspp4]{article}

\input{tcilatex}

\newcommand{\be}{\begin{equation}}
\newcommand{\ee}{\end{equation}}
\newcommand{\bea}{\begin{eqnarray}}
\newcommand{\eea}{\end{eqnarray}}

\input tcilatex
\begin{document}

\begin{titlepage}

\begin{center}
 
{\LARGE \bf Perturbed Friedmann Cosmologies Filled }
{\LARGE \bf with Dust and Radiation}
 
\vskip1cm
 
{\bf Z. Perj\'es}
\\
{\em KFKI Research Institute for Particle and Nuclear Physics, \\
H-1525 Budapest 114, P.O.Box 49, Hungary}
 
\today
 
\begin{abstract}

An infinite number of perturbed $k=0$ Friedmann
cosmologies filled with dust and radiation is found. As we
go up the sequence, the solutions contain an increasing
number of integration functions. With
the coordinate gauge adopted to co-move with the perturbed matter,
the solution of a pair of coupled equations
for the trace $h$ of the metric perturbations and for the radiation
density perturbations $\delta\rho_r$ is the key to the problem.
An estimate is given of the temperature variation of the cosmic microwave
background radiation due to the Sachs-Wolfe effect.
It is found that the temperature fluctuations can grow faster than
in the absence of radiation.
 
\end{abstract}
\end{center}
\end{titlepage}

\section{Introduction}

The first-order perturbations of a Friedmann universe with a flat 3-space
filled either with dust or radiation have been obtained by \cite{SW}. They
computed the temperature fluctuations of the cosmic background radiation in
the dust-filled universe, assuming that the domain in which the photon
travels is matter-dominated. This classic prediction for the temperature
fluctuations overestimates the experimental value (\cite{Ma}) by several
orders of magnitude. The prediction has been made by considering only the
gravitational perturbations along the path of the photon. Obviously, the
inclusion of random temperature fluctuations on the surface of the last
scattering can only increase the effect. Neither did the discovery of
large-scale structures (voids and walls) bring us closer to the resolution
of this paradox. Though it has been put forward (\cite{Ma}) that the
observed fluctuations are primordial (as opposed to propagation effects),
this is hard to take seriously unless we are able to reduce the magnitude of
the Sachs-Wolfe effect.

Here we compute the perturbations of the $k=0$ Friedmann universe in the
presence of {\em both} dust and radiation. Perturbations of two-fluid
cosmologies have been studied in the past decades (\cite{KS2,Koranda,Kom}).
Instabilities are known to exist in two-fluid cosmologies (\cite{Mukhanov}).
One could reduce the magnitude of the metric fluctuations affecting the
photon orbits by assuming, for example, that the observed large structures
are late emerging manifestations of the instabilities. The main result of
the present paper, in Sec. 6, is an infinite series of solutions, with an
increasing number of integration functions. Like in the pure dust model, the
solutions contain a relatively increasing and a relatively decreasing mode
of the radiation density perturbation. The growth is, however faster in the
presence of radiation and even faster as we go up the sequence of our
solutions. In Sec. 7, we estimate the magnitude of the Sachs-Wolfe effect on
the temperature variation of the cosmic background radiation. We find that
the relatively growing mode is incompatible with the observations already
with the first solution.

The energy-momentum tensor is a sum of those of the two media, 
\begin{equation}
T_{\;b}^a=T_{m\,b}^{\,a}+T_{r\,b}^{\,a}
\end{equation}
The contribution of the dust has the form 
\begin{equation}
T_{m\,b}^{\,a}=\rho _mu^au_b\ .  \label{Tm}
\end{equation}
For the radiation, 
\begin{equation}
T_{r\,b}^{\,a}={\tfrac 43}\rho _ru^au_b-{\tfrac 13}\rho _r\delta _{\;b}^a.
\label{Tr}
\end{equation}
The four-velocities are normalized $u^au_a=1$. We assume that, after the
decoupling, the conservation law $T_{b;a}^a\equiv $ $\left( \rho +p\right)
u_{a;b}u^b-p_{,a}+u_au^bp_{,b}=0$ applies separately to both the matter and
the radiation components: 
\begin{equation}
T_{mb;a}^a=0,\qquad \quad T_{rb;a}^a=0.
\end{equation}
This is justifiable because the decoupling occurs near the time of equal
matter-and-radiation density, and after that, coupling occurs only via the
universal gravitational interaction. We get two energy conservation
equations by transvecting with $u^b,$ 
\begin{equation}
\quad \left( \rho _mu_m^a\right) _{;a}=0,\qquad \quad \left( \rho
_r^{3/4}u_r^a\right) _{;a}=0.  \label{Econs}
\end{equation}
and two momentum conservation laws: 
\begin{equation}
u_{m\left[ a,b\right] }u_m^b=0,\qquad \quad 4\rho _r\left(
u_{ra,b}-u_{rb,a}\right) u^b=\rho _{r,a}-u_{ra}u_r^b\rho _{r,b}.
\label{momcons}
\end{equation}

\section{The perturbed model}

We adopt the conformal form of the metric 
\begin{equation}
g_{ab}=a^2(\eta )\left( \eta _{ab}+h_{ab}\right) ,  \label{g}
\end{equation}
where $\left( \eta _{ab}\right) =diag\left( 1,-1,-1,-1\right) $ and $h_{ab}$
is the metric perturbation. The indices of the perturbed quantities are
lowered and raised by the Minkowski metric $\eta _{ab}=\eta ^{ab}$. We shall
not use the explicit form of the scaling function $a(\eta )$ as long as
possible. In the perturbed universe, $h_{ab}\neq 0$, the densities of the
components can be written to first order

\begin{equation}
\rho _i^{(1)}=\rho _i+\delta \!\rho _i\ .
\end{equation}
where $i$ stands either for $m$ (matter) or $r$ (radiation). Here $\rho _r$
and $\rho _m$ are the unperturbed densities

\begin{equation}
\rho _m(\eta )=\rho _{m0}\frac{a_0^3}{a^3}\,,\qquad \qquad \rho _r(\eta
)=\rho _{r0}\frac{a_0^4}{a^4}
\end{equation}
and $\delta \rho _i$ are the first-order density perturbations. According to
the unperturbed models, the radiation density $\rho _r$ dies out faster than
the matter density, $\rho _m\ .$

The dipole effect of the cosmic background radiation provides an
experimental value for the local relative velocity of the order of $600$ $%
km/sec$. Thus we have good reason to assume that $\delta u_m^i\neq \delta
u_r^i$. Here we stick to a comoving gauge. For gauge invariant methods, {\it %
cf.} Refs. \cite{Mag,Russ,KS2,Dun,Mukhanov}. We choose coordinates comoving
with the matter: 
\begin{eqnarray}
u_m^a &=&u_0^a \\
u_r^a &=&u_0^a+\delta \!u_i^a
\end{eqnarray}
where the coincident unperturbed velocities are 
\begin{equation}
u_0^a={\frac 1a}\delta _0^a\ .
\end{equation}
The normalization conditions imply that $h_{00}=0$ and $\delta
\!u_r^0=\;\delta {u_r}_0\;=0$.

The permissible coordinate transformations in the comoving gauge are (\cite
{SW}) 
\[
\widehat{x}^a=x^a-\xi ^a 
\]
where the first-order function $\xi ^a$ has the form 
\begin{equation}
\xi ^0=\frac{b(x^\beta )}a,\qquad \xi ^\alpha =c^\alpha (x^\beta )
\label{gauge}
\end{equation}
and $x^\alpha $ are the space coordinates for $\alpha =1,2$ and $3.$ The
metric perturbations transform 
\begin{eqnarray}
\widehat{h}_{\alpha \beta } &=&h_{\alpha \beta }+c_{\alpha ,\beta }+c_{\beta
,\alpha }+2\frac{a^{\prime }}{a^2}b\eta _{\alpha \beta }  \label{hgauge} \\
\widehat{h}_{0\beta } &=&\frac{b_{,\beta }}a+h_{0\beta },\qquad \widehat{h}%
_{00}=0  \nonumber
\end{eqnarray}
where a prime ($^{\prime }$) denotes derivative with respect to the time
coordinate $\eta =x^0$. The velocity perturbation is gauge invariant to the
required order. The density perturbations transform as follows (\cite{Brauer}%
), 
\begin{equation}
\delta \!\hat{\rho}_i=\delta \!\rho _i+\frac ba\rho _i^{\prime }.
\label{rhogauge}
\end{equation}

\subsection{Energy conservation}

The energy conservation (\ref{Econs}) for the first-order perturbations of
the {\em dust} reads 
\begin{equation}
{\left( {\frac{\delta \!\rho _m}{\rho _m}}+{\frac 12}h\right) }^{\prime }=0
\end{equation}
where $h=h_\alpha ^\alpha $ is the trace of the metric perturbation and a
prime denotes partial derivative with respect to the conformal time $\eta $.
Hence we get

\begin{equation}
\delta \!\rho _m=\rho _m\;\left( E(x^\beta )-{\tfrac 12}h\right) .
\label{2d}
\end{equation}
Here the integration function$\;E(x^\beta )$ depends only on the space
coordinates $x^\alpha $ ($\alpha =1,2$ or $3$). A gauge transformation (\ref
{gauge}) alters $E$ as follows, 
\begin{equation}
\;\hat{E}(x^\beta )=\;E(x^\beta )+c_{,\alpha }^\alpha .  \label{Egauge}
\end{equation}

The energy conservation law (\ref{Econs}) for the {\em radiation} has the
form

\begin{equation}
{\left( \;\rho _r^{\frac 34}\,\sqrt{g}\,u_r^a\;\right) }_{,a}=0.
\end{equation}
Collecting the first-order terms, we have 
\begin{equation}
{\left( {\frac 34}{\frac{\delta \!\rho _r}{\rho _r}}+{\frac 12}h\right) }%
^{\prime }+a\left( \delta \!u_r^\alpha \right) _{,\alpha }=0.  \label{er}
\end{equation}

\subsection{Momentum conservation}

For the {\em dust,} the conservation law (\ref{momcons}) simplifies, 
\begin{equation}
\left( ah_{0\alpha }\right) ^{\prime }=0.
\end{equation}
This has the solution 
\begin{equation}
ah_{\alpha 0}=F_\alpha (x^\beta )\ .  \label{h0a}
\end{equation}

The coordinate freedom (\ref{gauge}) makes it possible to arrange (\cite
{White}) $F_{,\alpha }^\alpha =0,$ whence 
\begin{equation}
h_{\;\;\;,\alpha }^{0\alpha }=0\ .
\end{equation}
The remaining gauge transformations are still of the form (\ref{gauge}),
with 
\begin{equation}
\Delta b=0.  \label{remb}
\end{equation}
Here the Laplacian is defined by $\Delta b=-\eta ^{\alpha \beta }b,_{\alpha
\beta }.$

The momentum conservation law (\ref{momcons}) reads for the {\em radiation}
perturbations 
\begin{equation}
\left( {\rho _r}^{1/4}\delta {u_r}_\alpha \right) ^{\prime }={\tfrac 14}%
a\rho _r^{-3/4}\delta {\rho _r}_{,\alpha }\ .  \label{mr}
\end{equation}

\subsection{The potential $v$}

In (\cite{Kom}), the{\em \ }potential{\em \ }$v$ has been introduced under
the assumption that the vorticity of the radiation vanishes. We now show
that this potential exists for generic perturbations, without resorting to
any assumption about the vorticity. We introduce the potential $v=v\left(
\eta ,x^{\beta }\right) $ by writing 
\begin{equation}
{\tfrac{1}{4}}a\rho _{r}^{-3/4}\delta {\rho _{r}}=v^{\prime }{.}  \label{v}
\end{equation}
Under gauge transformations, 
\[
\hat{v}=v+\rho _{r}^{1/4}b. 
\]
Thus we can integrate Eq. (\ref{mr}) as follows: 
\begin{equation}
{\rho _{r}}^{1/4}\delta {u_{r}}_{\alpha }={v}_{,\alpha }  \label{va}
\end{equation}
such that the function of integration has no significance in (\ref{v}) and
is dropped. Hence the velocity perturbation $\delta {u_{r}}_{\alpha }=\delta
\left( g_{\alpha b}u_{r}^{b}\right) $ has the form 
\begin{equation}
\eta _{\alpha \beta }\delta \!u_{r}^{\beta }=\left[ \rho
_{r}^{-1/4}v_{,\alpha }\left( \eta ,x\right) -F_{\alpha }\left( x\right)
\right] a^{-2},  \label{deltaur}
\end{equation}
Substituting $\delta \!u_{r}^{\alpha }$ and $\delta {\rho _{r}}$ in the
perturbed energy conservation [Eq.(\ref{er})], we obtain 
\begin{equation}
{{3}v^{\prime \prime }-\Delta v+{\tfrac{1}{2}}\rho _{ro}^{1/4}a_{o}h}%
^{\prime }=0.  \label{veq}
\end{equation}

\section{Einstein equations}

The perturbed Einstein tensor will be written 
\begin{equation}
G_{\;b}^a={}_oG_{\;b}^a+\delta G_{\;b}^a\ .
\end{equation}
Here ${}_oG_{\;b}^a$ is the unperturbed tensor and $\delta G_{\;b}^a$ the
first-order part. The Einstein equations for the first-order quantities are 
\begin{eqnarray}
\delta G_{\;\;0}^0\; &=&-\left( \delta \!\rho _r+\delta \!\rho _m\right) \\
\delta G_{\;\;0}^\alpha \; &=&-{\tfrac 43}a\rho _r\delta \!u_r^\alpha \\
\delta G_{\;\;\beta }^\alpha \; &=&{\tfrac 13}\delta _{\;\,\beta }^\alpha
\delta \!\rho _r.
\end{eqnarray}
Substitution of the \cite{SW} expressions for $\delta G_{\;b}^a$ and
separating the trace-free part of the metric perturbation 
\begin{equation}
S_{\alpha \beta }=h_{\alpha \beta }-{\tfrac 13}\eta _{\alpha \beta }h
\end{equation}
yields 
\begin{equation}
S_{\quad ,\mu \nu }^{\mu \nu }+{\tfrac 23}\Delta h-2{\tfrac{a^{\prime }}a}%
h^{\prime }\,=\;-\,2a^2\left( \delta \!\rho _r+\delta \!\rho _m\right) ,
\label{el}
\end{equation}
\begin{equation}
S_{\quad ,\mu }^{\alpha \mu \quad \prime }-{\tfrac 23}{h^{,\alpha }}^{\prime
}+\Delta h^{0\alpha }-4\left( \,2\,\tfrac{{a^{\prime }}^2}{a^2}\,-\,\tfrac{%
a^{\prime \prime }}a\,\right) \;h^{0\alpha }\;={\tfrac 83}a^3\rho _r\delta
\!u_r^\alpha \;,  \label{ma}
\end{equation}
\begin{equation}
2h^{\prime \prime }+4\tfrac{a^{\prime }}ah^{\prime }-S_{\quad ,\mu \nu
}^{\mu \nu }-{\tfrac 23}\Delta h\;=\;-2a^2\delta \!\rho _r\ ,  \label{ha}
\end{equation}
\begin{eqnarray}
{S_{\;\beta }^\alpha }^{\prime \prime }+2\tfrac{a^{\prime }}a{S_{\;\beta
}^\alpha }^{\prime }-\Delta S_{\;\beta }^\alpha &=&S_{\quad ,\beta \mu
}^{\alpha \mu }+S_{\beta \mu ,}^{\quad \alpha \mu }-{\tfrac 23}\delta
_{\;\beta }^\alpha S_{\quad ,\mu \nu }^{\mu \nu }  \nonumber \\
&&+h_{\quad ,\beta }^{\alpha 0\quad \prime }+h_{\beta 0,}^{\quad \,\alpha
\,\prime }+2{\tfrac{a^{\prime }}a}\left( \>h_{\quad ,\beta }^{\alpha 0\quad
}+h_{\beta 0,}^{\quad \,\alpha }\>\right)  \nonumber \\
&&-{\tfrac 13}h,_{\;\beta }^\alpha -{\tfrac 19}\delta _{\;\beta }^\alpha
\Delta h.  \label{ne}
\end{eqnarray}

The sum of (\ref{el}) and (\ref{ha}) gives the simple relation

\begin{equation}
h^{\prime \prime }+\tfrac{a^{\prime }}ah^{\prime }=-a^2\left( 2\delta \rho
_r+\delta \!\rho _m\right) \ .  \label{7}
\end{equation}
Taking the divergence of Eq. (\ref{ma}) and subtracting the $\eta $
derivative of (\ref{el}), we get the integrability condition of these
equations. With the help of Eqs. (\ref{2d}), (\ref{7}) and (\ref{er}), we
may verify, however, that this integrability condition is satisfied
identically.

\section{Plan of solution}

In this section we review the steps of the solution procedure assuming no 
{\it a priori} knowledge of the scaling function $a$. First, we write the
radiation velocity perturbation $\delta {u_r}_\alpha =\delta \left(
g_{\alpha b}u_r^b\right) $ as 
\begin{equation}
\delta {u_r}_\alpha =\ a^2\eta _{\alpha \beta }\delta \!u_r^\beta
+ah_{0\alpha }.  \label{ua}
\end{equation}
Eliminating $\delta u_r^\alpha $ from Eqs (\ref{va}) and (\ref{deltaur}), we
obtain 
\begin{equation}
{\left( {\tfrac 32}{\tfrac{\delta \!\rho _r}{\rho _r}}+h\right) }^{\prime
\prime }-{\tfrac 12}\tfrac{\Delta \delta {\rho _r}}{{\rho _r}}=0.
\end{equation}
In the first place, the coupled system consisting of this and Eq. (\ref{7})
is solved for the radiation density perturbation $\delta \rho _r$ and for
the trace $h$. We next express $\delta \!u_r^\beta $ from Eqs. (\ref{mr})
and (\ref{ua}): 
\begin{equation}
\eta _{\alpha \beta }\delta \!u_r^\beta ={\tfrac 14a^{-2}\rho _r}%
^{-1/4}\tint a\rho _r^{-3/4}\delta {\rho _r}_{,\alpha }d\eta \
-a^{-1}h_{0\alpha }.
\end{equation}
We determine the divergences of the trace-free part from Eqs. (\ref{ha}) and
(\ref{ma}) as follows, 
\begin{eqnarray}
S_{\quad ,\mu \nu }^{\mu \nu }\; &=&2h^{\prime \prime }+4\tfrac{a^{\prime }}%
ah^{\prime }\;-{\tfrac 23}\Delta h+2a^2\delta \!\rho _r\   \label{Sdiv1} \\
S_{\quad ,\mu }^{\alpha \mu \quad } &=&{\tfrac 23}{h^{,\alpha }}-4\,\tfrac{%
a^{\prime }}a\,h^{0\alpha }-\tint \left( \Delta h^{0\alpha }+{\tfrac 83}%
a^2\rho _rh^{0\alpha }\right) d\eta  \nonumber \\
&&+{\tfrac 23}\tint a{\rho _r}^{3/4}\tint a\rho _r^{-3/4}\delta {\rho
_r^{\;,\alpha }}d\eta _1\ d\eta _2+S_{0\;,\mu }^{\alpha \mu \quad }.
\label{Sdiv2}
\end{eqnarray}
Here $h^{0\alpha }$ is given by Eq. (\ref{h0a}) such that $h^{0\alpha \prime
}=-\left( a^{\prime }/a\right) h^{0\alpha }$. The integration function $%
S_{0\;,\mu }^{\alpha \mu }=S_{0\;,\mu }^{\alpha \mu }\left( x^\beta \right)
\;$can be further determined by substituting $S_{\quad ,\mu }^{\alpha \mu
\quad }$ in Eq. (\ref{Sdiv1}). Upon inserting these expressions in Eq. (\ref
{ne}) we get the following inhomogeneous equation for $S_{\;\beta }^\alpha $%
: 
\begin{eqnarray}
{S_{\;\beta }^\alpha }^{\prime \prime }+2\tfrac{a^{\prime }}a{S_{\;\beta
}^\alpha }^{\prime }-\Delta S_{\;\beta }^\alpha &=&S_{0\;,\beta \mu
}^{\alpha \mu }+S_{0\beta \mu ,}^{\quad \;\;\alpha \mu }+{h_{\;\beta
}^{,\alpha }}+{\tfrac 13}\delta _{\;\beta }^\alpha \Delta h-3\,\tfrac{%
a^{\prime }}a\left( \>h_{\quad ,\beta }^{\alpha 0\quad }+h_{\beta 0,}^{\quad
\,\alpha }\>\right)  \nonumber \\
&&+{\tfrac 43}\tint a{\rho _r}^{3/4}\tint a\rho _r^{-3/4}\delta {\rho
_{r\;\beta }^{\;,\alpha }}d\eta _1\ d\eta _2  \nonumber \\
&&-\tint \left[ \Delta \left( \>h_{\quad ,\beta }^{\alpha 0\quad }+h_{\beta
0,}^{\quad \,\alpha }\>\right) +{\tfrac 83}a^2\rho _r\left( \>h_{\quad
,\beta }^{\alpha 0\quad }+h_{\beta 0,}^{\quad \,\alpha }\>\right) \right]
d\eta  \nonumber \\
&&-{\tfrac 43}\delta _{\;\beta }^\alpha \left( h^{\prime \prime }+2\tfrac{%
a^{\prime }}ah^{\prime }+a^2\delta \!\rho _r\right) .  \label{ne2}
\end{eqnarray}

\section{k=0 universes}

The unperturbed ($h_{ab}=0$) metric satisfies (\cite{MTW}) 
\begin{equation}
3{\frac 1{a^2}}{\left( {\frac{da}{d\eta }}\right) }^2-\frac{{\rho _m}_o{a_o}%
^3}a-\frac{{\rho _r}_o{a_o}^4}{a^2}=0\ 
\end{equation}
such that the density $\rho _i$ equals ${\rho _m}_o$ or ${\rho _m}_r$ at
some prescribed conformal time $\eta =\eta _0$. Using the Hubble constant 
\[
H=\frac{24}{\rho _{m_o}a_0^3} 
\]
(which is realistic in a matter-dominated ambience at $\eta =\eta _0$), this
equation can be written 
\[
{\left( {\frac{da}{d\eta }}\right) }^2-\frac 8H\left( a+\frac 2H\mu
^2\right) =0 
\]
where 
\[
\mu ^2=\frac{{\rho _r}_oa_o}{2\rho _{m_o}}H 
\]
is a constant.

Hence, fixing the origin of $\eta ,$ the solution has the form 
\begin{equation}
a=\tfrac 2H\left( \eta ^2-\mu ^2\right) .
\end{equation}
Normalizing the time coordinate by 
\[
\xi =\frac \eta \mu , 
\]
we have 
\begin{equation}
a=\tfrac 2H\mu ^2\left( \xi ^2-1\right)
\end{equation}
so that the Big Bang occurs at $\xi =1$. The range of the time coordinate is
determined by the ratio of the radiation density to matter density\footnote{%
Small values, $\left| \xi \right| <1,$ are compatible with a big crunch.},
thus in our Universe, $\xi >1$.

The evolution of the trace $h$, Eq. (\ref{7}), is driven by the velocity
potential [Eq. (\ref{v})]: 
\begin{equation}
h^{\prime \prime }+\frac{a^{\prime }}ah^{\prime }+\frac{12}{Ha}\left(
2E(x^\beta )-h\right) =-8a{\rho _r}^{3/4}v^{\prime }  \label{f0}
\end{equation}
where Eq. (\ref{2d}) was used for the matter density perturbation. We may
get rid of the inhomogeneous term $E(x^\beta )$ by introducing the function 
\begin{equation}
f=h-2E(x^\beta ).
\end{equation}
Gauge transformations alter $f$ as follows, 
\begin{equation}
\hat{f}=f+6\tfrac{a^{\prime }}{a^2}b.  \label{fgauge}
\end{equation}
With the new time variable $\xi $, Eq. (\ref{f0}) takes the form 
\begin{equation}
\ddot{f}+\tfrac{2\xi }{\xi ^2-1}\dot{f}-\tfrac 6{\xi ^2-1}{f}=-\tfrac{16}{%
\left( \xi ^2-1\right) ^2}K\dot{v}  \label{fi}
\end{equation}
where an overdot means $d/d\xi $ and 
\[
K=\sqrt{3}\rho _{m_o}{\rho _r}_o^{-3/4}. 
\]
Equation (\ref{veq}) becomes 
\begin{equation}
{\ddot{v}-}\tfrac{{\mu ^2}}{{3}}{\Delta v+{\tfrac 1K}\dot{f}}=0.
\label{veq2}
\end{equation}

We now introduce a new, {\em gauge-invariant} potential $u$ by writing 
\begin{equation}
u=K\dot{v}+f.  \label{udef}
\end{equation}
Equation (\ref{fi}) and the time derivative of Eq. (\ref{veq2}) then are,
respectively, 
\begin{mathletters}
\begin{eqnarray}
\ddot{f}+\tfrac{2\xi }{\xi ^2-1}\dot{f}-\left( \tfrac 6{\xi ^2-1}+\tfrac{16}{%
\left( \xi ^2-1\right) ^2}\;\right) {f} &=&-\tfrac{16}{\left( \xi
^2-1\right) ^2}u  \label{fi2} \\
\ddot{u}-\tfrac{{\mu ^2}}3{\Delta }\left( {u-f}\right) &=&0{.}  \label{ueq}
\end{eqnarray}
From (\ref{ueq}), we can express $\Delta f$ and substitute it in the
equation obtained by acting with the Laplacian on Eq. (\ref{fi2}). Thus we
get the fourth-order equation for $u:$%
\end{mathletters}
\begin{equation}
\left( \tfrac{d^2}{d\xi ^2}+\tfrac{2\xi }{\xi ^2-1}\tfrac d{d\xi }-\tfrac
6{\xi ^2-1}\right) \Delta u-\tfrac 3{{\mu }^2}\left( \tfrac{d^2}{d\xi ^2}+%
\tfrac{2\xi }{\xi ^2-1}\tfrac d{d\xi }-\tfrac 6{\xi ^2-1}-\tfrac{16}{\left(
\xi ^2-1\right) ^2}\right) \tfrac{d^2u}{d\xi ^2}=0.  \label{fourth}
\end{equation}
Given a solution of the coupled equations (\ref{fi2}) and (\ref{ueq}), the
radiation density perturbation can be computed from Eq. (\ref{v}) as
follows: 
\begin{equation}
\delta {\rho _r}=\tfrac 4{\mu a}\rho _r^{3/4}\dot{v}=\tfrac 4{\mu K}\rho
_r^{3/4}\tfrac{u-f}a.  \label{derhsol}
\end{equation}

\section{Particular solutions}

\label{a:eqh.mp}

$(i)$ First we seek solutions with a vanishing gauge-invariant potential, $%
u=0.$ By Eq. (\ref{ueq}), then $f$ is a harmonic function, $\Delta f=0,$ and
Eq. (\ref{fi2}) becomes 
\begin{equation}
\ddot{f}+\frac{2\xi }{\xi ^2-1}\dot{f}-\left( \frac 6{\xi ^2-1}+\frac{16}{%
\left( \xi ^2-1\right) ^2}\right) f=0.  \label{genLeg}
\end{equation}
This is the generalized Legendre equation for $\nu =2$ and $n=4$. Particular
solutions are 
\[
P_2^4(\xi )=Q_2^4(\xi )\int^\xi \frac 1{\left( \zeta ^2-1\right) \left[
Q_2^4(\zeta )\right] ^2}d\zeta 
\]
and the associated Legendre function of the second kind 
\begin{equation}
Q_2^4(\xi )=\left( 1-\xi ^2\right) ^2\frac{d^4Q_2}{d\xi ^4}.
\end{equation}
Inserting here the Legendre functions $P_2=\frac 12\left( 3\xi ^2-1\right) $
and $Q_2=\frac 12P_2\ln \frac{\xi +1}{\xi -1}-\tfrac 32\xi $, we get the
particular solutions in the form 
\begin{eqnarray}
f_1 &\equiv &48P_2^4=\frac 15\frac{\xi ^6-5\xi ^4+15\xi ^2+5}{\left( \xi
^2-1\right) ^2}  \label{fpart} \\
f_2 &\equiv &\tfrac 1{48}Q_2^4=\frac \xi {\left( \xi ^2-1\right) ^2}. 
\nonumber
\end{eqnarray}
Thus the solution of Eq.\ (\ref{genLeg}) is 
\begin{equation}
f=A_1\left( x^\alpha \right) f_1+A_2\left( x^\alpha \right) f_2
\label{fsoli}
\end{equation}
where the combination functions $A_1$ and $A_2$ are harmonic, $\Delta
A_1=\Delta A_2=0$, to yield the property $\Delta f=0$ as required. We may
then perform a gauge transformation (\ref{fgauge}) with a harmonic $b$ such
that $A_2=0$ is set in the solution (\ref{fsoli}).

$(ii)$ Our second set of particular solutions arises from the assumption
that $\Delta u=0.$ In this case we see from Eq. (\ref{fourth}) that it is $%
\ddot{u},$ rather than $f,$ that satisfies the generalized Legendre equation
(\ref{genLeg}). Hence 
\begin{equation}
u=A_1\left( x^\alpha \right) u_1+A_2\left( x^\alpha \right) u_2+C\left(
x^\alpha \right) \xi +D\left( x^\alpha \right)  \label{u}
\end{equation}
where the two independent solutions are 
\begin{eqnarray*}
u_1 &\equiv &48\tiint P_2^4d\xi _1d\xi _2=\tfrac 1{60}\xi ^4-\tfrac 3{10}\xi
^2-\tfrac 45\ln (\xi ^2-1) \\
u_2 &\equiv &\tfrac 1{48}\tiint Q_2^4d\xi _1d\xi _2=\tfrac 14\ln \tfrac{\xi
+1}{\xi -1}.
\end{eqnarray*}
It follows from $\Delta u=0$ that each of the integration functions $%
A_1\left( x^\alpha \right) ,$ $A_2\left( x^\alpha \right) ,$ $C\left(
x^\alpha \right) $ and $D\left( x^\alpha \right) $ is harmonic.

A particular solution of Eq. (\ref{fi2}) is given by (\cite{Bronstein}) 
\[
f_0(\xi )=16\int \frac{u\left( \zeta \right) }{\zeta ^2-1}\left[ f_2(\xi
)f_1(\zeta )-f_1(\xi )f_2(\zeta )\right] d\zeta .
\]
Since $u$ is a harmonic function, so is $f_0.$ Using (\ref{fpart}), we
obtain 
\begin{eqnarray}
f_0(\xi ) &=&\tfrac 15\tfrac 1{(\xi ^2-1)^2}\left\{ \left[ -\tfrac
4{15}\left( \xi ^2-3\right) \xi A_1+\tfrac 18\left( 3\xi ^4-6\xi
^2-25\right) A_2+\left( \xi ^4-2\xi ^2+5\right) C\right] \right.   \nonumber
\\
&&\qquad \qquad \qquad \times \left[ \left( \xi ^2+1\right) \ln \tfrac{\xi -1%
}{\xi +1}-2\xi \ln \left( \xi ^2-1\right) \right]   \nonumber \\
&&\qquad \qquad +\left[ -\tfrac 2{15}\left( \xi ^4-2\xi ^2+125\right)
A_1+\tfrac 14\left( 3\xi ^2-9\right) \xi A_2+\left( 2\xi ^2-6\right) \xi
C\right]   \label{f0sol} \\
&&\qquad \qquad \qquad \times \left[ \left( \xi ^2+1\right) \ln \left( \xi
^2-1\right) -2\xi \ln \tfrac{\xi -1}{\xi +1}\right]   \nonumber \\
&&\qquad \qquad -\left( 3-\tfrac{213}5\xi ^2-\tfrac 4{75}\xi ^6+\tfrac{176}{%
45}\xi ^4\right) A_1+\tfrac 14\left( 7-14\xi ^2+3\xi ^4\right) \xi A_2 
\nonumber \\
&&\qquad \qquad \left. +2(2\xi ^2+\xi ^4+5)\xi C+20\left( 1+\xi ^2\right)
D\right\} .  \nonumber
\end{eqnarray}

The solution of the inhomogeneous Eq. (\ref{fi2}) has the form 
\begin{equation}
f=f_0+F_1\left( x^\alpha \right) f_1+F_2\left( x^\alpha \right) f_2.
\label{fsolii}
\end{equation}
Equation (\ref{ueq}) has yet to be satisfied: 
\[
\ddot{u}+\tfrac{{\mu ^2}}3\Delta f=0. 
\]
Inserting here (\ref{u}), we get 
\[
A_1\left( x^\alpha \right) f_1+A_2\left( x^\alpha \right) f_2+\tfrac{{\mu ^2}%
}3\Delta f=0. 
\]
Hence 
\[
A_1=\tfrac{{\mu ^2}}3\Delta F_1,\qquad A_2=\tfrac{{\mu ^2}}3\Delta F_2 
\]
where $F_1$ and $F_2$ are biharmonic functions, {\em i.e.}, $\Delta \Delta
F_1=\Delta \Delta F_2=0.$

$(iii)$ We may continue the process of generating new solutions by replacing
next the harmonic condition on $u$ with a biharmonic condition. Thus the
function $\Delta \Delta f$ satisfies the homogeneous equation (\ref{genLeg}%
). The solution of this will define $\Delta u$ by the equation 
\begin{equation}
\Delta \ddot{u}=-\tfrac{{\mu ^2}}3{\Delta \Delta f.}  \label{lapueq}
\end{equation}
We take the Laplacian of both sides of Eq. (\ref{fi2}) and solve for $\Delta
f$, given the source term $\Delta u.$ This in turn yields the potential $u$
by using Eq. (\ref{ueq}). A comparison of (\ref{fsoli}) and (\ref{fsolii})
reveals how the solution generating procedure essentially proceeds: the next
solution in the sequence is generated by relaxing the harmonic condition on
the coefficients of the given solution $f$.

The general solution of Eq. (\ref{f0}) is 
\begin{equation}
h=f_0(\xi )+A_1f_1(\xi )+A_2f_2(\xi )+2E(x^\beta ).  \label{hsol}
\end{equation}
where $A_1=A_1(x^\beta )$ and $A_2=A_2(x^\beta )$ are integration functions$%
. $ The gauge transformations (\ref{hgauge}) with a nonvanishing $c^\alpha $
parameter alter $h$ as follows: $\hat{h}=h+2c_{,\alpha }^\alpha .$ Thus, by
a suitable gauge transformation, we arrange that $E=0,$ and still we can
perform transformations with $c_{,\alpha }^\alpha =0.$

Finally, we get the trace-free part ${S_{\;\beta }^\alpha }$ of the metric
perturbation from Eq. (\ref{ne}). The solution will have the form 
\[
{S_{\;\beta }^\alpha =}\stackunder{0}{S}{_{\;\beta }^\alpha +}\stackunder{1}{%
S}{_{\;\beta }^\alpha } 
\]
where $\stackunder{0}{S}{_{\;\beta }^\alpha }$ is a {\it spheroidal wave
function }(\cite{Flammer,Stratton,Fisher}), a solution of the homogeneous
equation 
\begin{equation}
{\ddot{S}_{\;\beta }^\alpha }+\tfrac{4\xi }{\xi ^2-1}{\dot{S}_{\;\beta
}^\alpha }-\mu ^2\Delta S_{\;\beta }^\alpha =0  \label{Shom}
\end{equation}
and $\stackunder{1}{S}{_{\;\beta }^\alpha }$ is a particular solution of (%
\ref{ne}).

Let us consider the solutions which are given by $C^\infty $ functions.
Following \cite{White}, we may then represent the amplitude $A_1$ in terms
of a $C^\infty $ function $B$ as follows, 
\begin{equation}
A_1=\Delta B.
\end{equation}
The treatment of the curl terms containing $h^{0\alpha }$ is a fairly
straightforward task. There remain to be found the pure density
perturbations with $h^{0\alpha }=\stackunder{0}{S}{_{\;\beta }^\alpha =}0.$
For these perturbations, Eq. (\ref{ne}) simplifies somewhat, 
\begin{eqnarray}
{\ddot{S}_{\;\beta }^\alpha }+\tfrac{4\xi }{\xi ^2-1}{\dot{S}_{\;\beta
}^\alpha }-\mu ^2\Delta S_{\;\beta }^\alpha &=&\mu ^2\left( S_{0\;,\beta \mu
}^{\alpha \mu }+S_{0\beta \mu ,}^{\quad \;\;\alpha \mu }+{h_{\;\beta
}^{,\alpha }}+{\tfrac 13}\delta _{\;\beta }^\alpha \Delta h\right)  \nonumber
\\
&&-{\tfrac 43}\delta _{\;\beta }^\alpha \left( \ddot{h}+\tfrac{4\xi }{\xi
^2-1}\dot{h}+\mu ^2a^2\delta \!\rho _r\right)  \nonumber \\
&&+{\tfrac 43}\mu ^4\tint a{\rho _r}^{3/4}\tint a\rho _r^{-3/4}\delta {\rho
_{r\;\beta }^{\;,\alpha }}d\xi _1\ d\xi _2.  \label{ne3}
\end{eqnarray}
For solution $(i)$, we have $h=\Delta Bf_1$ and $\delta {\rho _r}=-\tfrac
4{\mu Ka}\rho _r^{3/4}h,$ whence 
\begin{eqnarray}
{\ddot{S}_{\;\beta }^\alpha }+\tfrac{4\xi }{\xi ^2-1}{\dot{S}_{\;\beta
}^\alpha }-\mu ^2\Delta S_{\;\beta }^\alpha &=&\mu ^2\left( S_{0\;,\beta \mu
}^{\alpha \mu }+S_{0\beta \mu ,}^{\quad \;\;\alpha \mu }+{h_{\;\beta
}^{,\alpha }}-{\tfrac{32}3}\Delta B{_{\;\beta }^{,\alpha }}\tint \tfrac{{1}}{%
\left( \xi ^2-1\right) ^2}\tint f_1d\xi _1\ d\xi _2\right)  \nonumber \\
&&-{\tfrac 83}\delta _{\;\beta }^\alpha \Delta B.  \label{nei}
\end{eqnarray}
We seek a particular solution in the form 
\begin{equation}
\stackunder{1}{S}{_{\;\beta }^\alpha }=\Delta B_{\;\beta }^{,\alpha }F\left(
\xi \right) -\left( B_{\;\beta }^{,\alpha }+\tfrac 13\delta _\beta ^\alpha
\Delta B\right) X\left( \xi \right) +S_{0\beta }^\alpha  \label{S1ab}
\end{equation}
where the requirement of compatibility with the divergence equation (\ref
{Sdiv2}) yields 
\begin{eqnarray}
X\left( \xi \right) &=&f_1-{8}\tint \tfrac{{1}}{\left( \xi ^2-1\right) ^2}%
\tint f_1d\xi _1\ d\xi _2 \\
&=&\tfrac 1{15}\left[ \tfrac{3\xi ^4-12\xi ^2+1}{\xi ^2-1}-4\ln (\xi
^2-1)\right]  \nonumber
\end{eqnarray}
and $F\left( \xi \right) $ is a function to be determined. By substituting (%
\ref{S1ab}) into (\ref{nei}), we get the ordinary differential equation 
\begin{equation}
{\ddot{F}_{\;\beta }^\alpha }+\tfrac{4\xi }{\xi ^2-1}{\dot{F}_{\;\beta
}^\alpha }=-{\tfrac 83}\mu ^2\tint \tfrac{{1}}{\left( \xi ^2-1\right) ^2}%
\tint f_1d\xi _1\ d\xi _2.
\end{equation}
This has the solution 
\begin{eqnarray}
F &=&-\tfrac 2{10125}\tfrac{\mu ^2}{\xi ^2-1}\left[ -63\xi ^4-1249\xi
^2+\left( 45\xi ^4+544\xi ^2-709\right) \ln (\xi ^2-1)\right.  \nonumber \\
&&\left. +120\xi \ln \tfrac{\xi -1}{\xi +1}-120(\xi ^2-1)\ln (\xi -1)\ln
(\xi +1)\right] \\
&&+c_1+c_2\left( 2\tfrac \xi {\xi ^2-1}+\ln \tfrac{\xi -1}{\xi +1}\right) 
\nonumber
\end{eqnarray}
where $c_1$ and $c_2$ are integration constants. The terms in $\stackunder{1%
}{S}{_{\;\beta }^\alpha }$ proportional to $c_2$ have the harmonic amplitude 
$\Delta B_{\;\beta }^{,\alpha }.$ These terms solve the homogeneous
equation. They are a special case of gravitational waves described by
spheroidal wave functions. This reflects on the ambiguous nature of the
decomposition of the perturbations into wave and non-wave parts.

Collecting the results, the tensor perturbation of solution $(i)$ has the
form 
\begin{equation}
h{_{\alpha \beta }}=\stackunder{1}{S}{_{\alpha \beta }}+\tfrac{1}{3}\eta {%
_{\alpha \beta }}\Delta Bf_{1}  \label{hi}
\end{equation}
where $\stackunder{1}{S}{_{\alpha \beta }}$ and $f_{1}$ are given in Eqs. (%
\ref{S1ab}) and (\ref{fpart}), respectively.

\section{ The Sachs-Wolfe effect}

\label{a:eqs2.mp}

The temperature variation $\delta T$ of the cosmic background radiation can
be computed (\cite{SW}) as follows, 
\begin{equation}
\frac{\delta T}T=\frac 12\int_0^{\eta _R-\eta _E}\left( \frac{\partial
h_{\alpha \beta }}{\partial \eta }e^\alpha e^\beta -2\frac{\partial
h_{0\beta }}{\partial \eta }e^\beta \right) dw  \label{deTT1}
\end{equation}
where $\eta _R$ and $\eta _E$ denote the time of reception and emission,
respectively, and $w$ is the affine length along the null geodesic of
propagation with tangent 
\begin{equation}
\frac{dx^a}{dw}=\left( -1,e^\alpha \right)
\end{equation}
such that $e^\alpha e_\alpha =-1.$ We consider the contribution of the
relatively increasing mode. Then $h_{0\beta }=0$ and the second term under
the integral in Eq. (\ref{deTT1}) vanishes. The term $\stackunder{1}{S}{%
_{\alpha \beta }}$ has the amplitude $B_{,\alpha \beta }.$ By using the
relation 
\begin{equation}
y_{,a}\frac{dx^a}{dw}dw=y_{,\alpha }e^\alpha dw-y^{\prime }d\eta ,
\end{equation}
we get dipole anisotropy contributions with respective amplitudes $B_{,\beta
}e^\beta $ and $\Delta B_{,\beta }e^\beta $ and gravitational redshift
terms. The trace part of $h_{\alpha \beta }$, unlike that of the pure dust,
is time-dependent, and thus the cancellation of the integrated terms in the (%
\cite{SW}) result does not occur here. Taken together and excluding dipole
anisotropies, the contributions to the temperature variation sum up to 
\begin{equation}
\frac{\delta T}T=\tau _R-\tau _E+\Delta \tau  \label{deTT}
\end{equation}
where $\tau _R$ and $\tau _E$ are the values of the function 
\begin{eqnarray}
\tau &=&-\tfrac 2{675}\tfrac{\Delta B}{(\xi +1)^3}(3\xi ^2+9\xi ^2+13\xi
+15)\ln \left( \xi -1\right) \\
&&-\tfrac 2{675}\tfrac{\Delta B}{(\xi -1)^3}(3\xi ^2-9\xi ^2+13\xi -15)\ln
\left( \xi +1\right)  \nonumber \\
&&-\tfrac 1{15}\tfrac B{(\xi ^2-1)^3}(3\xi ^6-5\xi ^4-15\xi ^2-15)  \nonumber
\\
&&+8\tfrac{\Delta B}{(\xi ^2-1)^3}\left[ c_2\xi -\tfrac 1{10125}(18\xi
^6-220\xi ^4-885\xi ^2-225)\right]  \nonumber
\end{eqnarray}
at the respective events $R$ of reception and $E$ of emission. In addition
to these contributions representing the original Sachs-Wolfe effect, we have
the integrated term 
\begin{equation}
\Delta \tau =\int\limits_0^{\eta _R-\eta _E}\left[ \Delta B\tfrac \partial
{\partial \eta }\left( \tfrac{\partial ^2}{\partial \eta ^2}F-\tfrac 83%
\tfrac{{1}}{\left( \xi ^2-1\right) ^2}\tint f_1d\xi \right) -B\tfrac{%
\partial ^3}{\partial \eta ^3}X\right] dw.
\end{equation}
At large values of $\eta =\mu \xi ,$ the function $\tau \ $increases
logarithmically$.$ The integrated term $\Delta \tau $ gives a negligible
contribution at late times, but it must be taken into account in the
vicinity of $\xi =1.$ This indicates that the temperature fluctuations in
solution $(i)$ can be larger than for a pure dust cosmology.

\section{Discussion of the results}

The evolution of density contrast of the incoherent matter, $\delta \!\rho
_m/\rho _m$ mirrors the evolution of the trace $h,$ as can be seen from Eq. (%
\ref{2d}). For late times, that is in a matter-dominated era, one should
reasonably expect this density contrast to be well-approximated by the
Sachs-Wolfe scenario. However, for solution $(i)$, the trace perturbation
with the coefficient $A_1$ behaves asymptotically as $f_1\propto \xi ^2,$
unlike the relatively growing mode of \cite{SW}. The perturbation with the
coefficient $A_2$ is $h_2\propto 1/\xi ^3,$ precisely as the relatively
decreasing mode of \cite{SW}. The trace-free part of solution $(i)$ at late
times diverges also faster than in the absence of radiation: it goes like $%
h_1\propto \xi ^2\ln \xi .$ The asymptotic behavior of solution $(ii)$ is
similar: the coefficients of $A_2$ and $C$ tend to stationary values, and
the coefficient of $D$ is asymptotically $\propto 1/\xi ^2.$

One may ask whether or not the perturbative cosmologies obtained here are
favored by the exact, nonlinear evolutionary processes. This issue may be
addressed by numerical methods, to be discussed elsewhere (\cite{Czinner}).

{\bf Acknowledgement}

I thank M. Vas\'{u}th for illuminating discussions. This work has been
supported by the {\sl OTKA} fund T031724.

\end{document}